\documentclass[
final,
3p,
times,
number,
11pt,
sort&compres
]{elsarticle}

\journal{ArXiv.org}

\usepackage{amsmath}
\usepackage{amssymb}
\usepackage{latexsym}

\usepackage{booktabs}

\usepackage{graphicx}
\usepackage{subfig}

\usepackage{calc}
\usepackage{hyperref}
\hypersetup{
	colorlinks,
	linkcolor=blue,
	citecolor=cyan
}

\providecommand{\ud}{\mathrm{d}}
\newcommand{\degree}{^\circ}
\newcommand{\TT}{$\times$}
\providecommand{\ep}{$e^{\scriptscriptstyle-}\!/\!p^{\scriptscriptstyle+}$}
\newcommand{\RB}{\textbf{R}$\times$\textbf{B}{} }

\begin{document}

\bibliographystyle{model1a-num-names} 

\begin{frontmatter}

\title{\textbf{R}$\times$\textbf{B} Drift Momentum Spectrometer}

\author[wien]{X.~Wang}

\author[wien]{G.~Konrad}

\author[wien]{H.~Abele\corref{cor1}}
\ead{abele@ati.ac.at}

\cortext[cor1]{Corresponding author.}

\address[wien]{Technische Universit\"{a}t Wien, Atominstitut, Stadionallee 2, 1020 Vienna, Austria}

\begin{abstract}

We propose a new type of momentum spectrometer, which uses the \textbf{R}$\times$\textbf{B} drift effect to disperse the charged particles in a uniformly curved magnetic field. This kind of \textbf{R}$\times$\textbf{B} spectrometer is designed for the momentum analyses of the decay electrons and protons in the PERC (Proton and Electron Radiation Channel) beam station, which provides a strong magnetic field to guide the charged particles in the instrument. Instead of eliminating the guiding field, the \textbf{R}$\times$\textbf{B} spectrometer evolves the field gradually to the analysing field, and the charged particles can be adiabatically transported during the dispersion and detection. The drifts of the particles have similar properties as their dispersion in the normal magnetic spectrometer. Besides, the \textbf{R}$\times$\textbf{B} spectrometer is especially ideal for the measurements of particles with low momenta and relative large incident angles. We present a design of the \textbf{R}$\times$\textbf{B} spectrometer, which can be used in PERC. The resolution of the momentum spectra can reach 14.4 keV/c, if the particle position measurements have a resolution of 1 mm.

\end{abstract}


\end{frontmatter}

\section{Introduction}

The beam station PERC (Proton and Electron Radiation Channel) \cite{Dub08}, for the experiments of free neutron $\beta$ decay, is under development \cite{Kon12}. The motivation of PERC is to supply an intense beam of well defined electrons and protons (\ep) from free neutron decay. With the general-purpose \ep-beam, various quantities related to the physics in and beyond the Standard Model can be measured \cite{Jac57,Her01,Sev06,Abe08,Dub11}. 

\begin{figure}[htb]
	\centering
	\includegraphics[width=0.5\textwidth]{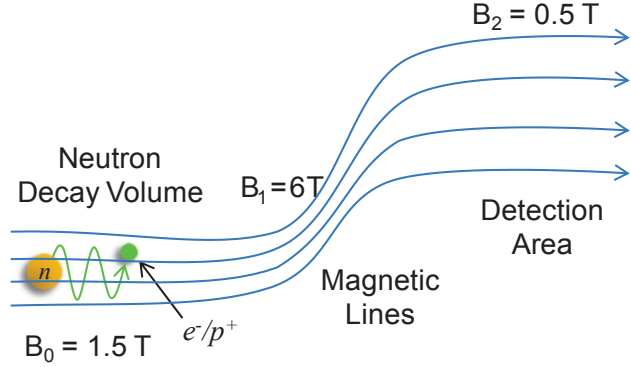}
	\caption{The principle of the PERC beam station. The charged decay \ep are guided by a magnetic field from the decay volume to the detection area.	}
	\label{figure:perc sketch}
\end{figure}

PERC provides in the instrument a neutron decay volume, where the free neutrons are imported and decay into charged electrons, protons, and neutral electron anti-neutrinos. With a series of coils, a specified static magnetic field is applied in the instrument. The charged decay \ep spiral along the magnetic field lines, and are guided from the decay to the detection area. Fig.~\ref{figure:perc sketch} sketches the principle of PERC. 

The neutron decay volume has a magnetic field of $B_0$ = 1.5 T. After the decay volume, a magnetic field barrier $B_1$ = 6 T is applied. The \ep particles propagate adiabatically in PERC, hence the pitch angles of them, i.e., the angles between the particle momentum $\mathbf{p}$ and the magnetic field $\mathbf{B}$, fulfil 

\begin{equation}
	\frac{\sin\theta_0}{\sin\theta_1} = \sqrt{\frac{B_0}{B_1}}
	\label{form:adia theta}
\end{equation} 

Therefore, only the \ep with pitch angles at $B_0$ smaller than the critical angle $\theta_c$ can pass through the $B_1$ field barrier

\begin{equation}
	\theta_0 \leqslant \theta_c = \arcsin ( \sqrt{\frac{B_0}{B_1}} \cdot \sin 90\degree) = 30\degree
\end{equation}

After the $B_1$ barrier, the guiding field is gradually decreased to $B_2$ = 0.5 T at the detection area, where the \ep particles can be processed and measured.

\begin{figure}[htb]
	\centering
	\includegraphics[width=0.45\textwidth]{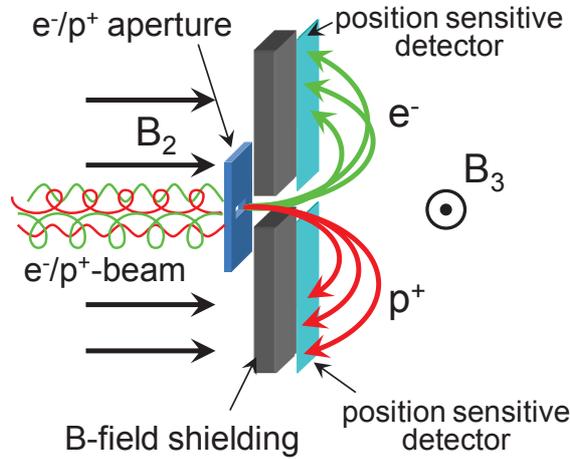}
	\caption{The principle of a magnetic spectrometer, that is attached to the exit of PERC.}
	\label{figure:principle mag_spec}
\end{figure}

As for the measurements of the particles, the loss free energy spectroscopy for electrons has been demonstrated in \cite{Bop86, Abe93}. While in the experiments with PERC, a spectrometer for the \ep momentum measurements besides the energy sensitive detector is desired. The principle of a normal magnetic spectrometer after PERC is sketched in Fig.~\ref{figure:principle mag_spec}. 

At the end of PERC, the magnetic spectrometer must firstly shield the guiding field $B_2$, and apply a vertical analysing magnetic field $B_3$. The incident \ep from PERC pass through a small aperture, then disperse in the $B_3$ field. The position sensitive detectors for electrons and protons are placed on both sides of the incident window. The dispersion distance of a particle in the spectrometer is then

\begin{equation}
	D(p,\theta) = 2 \frac{p}{qB}\cdot f(\theta) , \quad \mathrm{with} \quad f(\theta) = \cos \theta
	\label{form:magspec disperse}
\end{equation}

where $p$ and $q$ are the momentum and charge of the particle, $\theta$ is its incident angle according to the normal of the detector plane.

Compared with the energy resolving detectors, the magnetic spectrometer is versatile, that it can realize various measurements in PERC \cite{Dub08}, also detect the electrons and protons at the same time. Technically, the position sensitive detectors highly suppress the backscattering problem of \ep and the $\gamma$ background. The momentum spectrum of the decay electrons has higher resolution in the low energy scale, thus is especially needed for the estimation of the Fierz interference term $b_F$ in neutron decay \cite{Jac57}.

\begin{figure}[htb]
	\centering
	\includegraphics[width=0.4\textwidth]{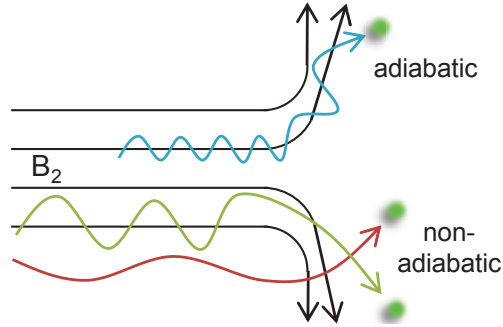}
	\caption{Sketch of \ep motions in the drastically decreased $B_2$ field. In adiabatic transports, the particles follow the magnetic field lines to vertical directions. While in non-adiabatic transports, the \ep are bended in the vertical field. In both cases the pitch angles of the particles are highly distorted.}
	\label{figure:magspec fail}
\end{figure}

However, because of the strong magnetic guiding field of PERC, we found difficulties in the design of the magnetic spectrometer. The spectrometer has to drastically decrease the guiding field $B_2$ to zero at the incident window, whereas the magnetic lines do not vanish, but spread in vertical directions. When the \ep pass the vertical field, the pitch angles of them are highly distorted either in adiabatic or in non-adiabatic transports, as shown in Fig.~\ref{figure:magspec fail}.

The distortion of the \ep pitch angles strongly depends on the field distribution and the \ep momenta, thus are not predictable nor controllable \cite{Var71}. Therefore, the distribution of the particles in the spectrometer can hardly represent their momenta. 

In this case, we propose a method of \RB drift momentum spectrometer, which can realize the momentum analyses of the \ep without eliminating the guiding field of PERC.

In Section \ref{section:principle}, we introduce the principle of the \RB spectrometer. The systematic corrections of the particle distribution are stated in Section \ref{section:correction}. The transfer function of the particles is discussed in Section \ref{section:TF}.

\section{Principle of the \RB drift momentum spectrometer}
\label{section:principle}

When a charged particle propagates in a curved magnetic field, it has the drift effect perpendicular to the magnetic field $\mathbf{B}$ and the field curvature $\mathbf{R}$, so called as the \RB drift. The drift velocity $\mathbf{v}_d$ of the first order can be expressed as \cite{Din05}

\begin{equation}
	\mathbf{v}_d = \frac{m}{qRB} \left(v^2_\parallel + \frac{1}{2}v^2_\bot\right) \frac{\mathbf{R}\times\mathbf{B}}{RB}
	\label{form:drift velocity}
\end{equation}

where $m$ is the mass of the particle, $v_\parallel$ and $v_\bot$ are the velocity components parallel and vertical to the magnetic field line. In a static magnetic field, the velocity components can be expressed with the absolute velocity $v$ and the pitch angle $\theta$

\begin{equation}
	 v_\parallel = v\cdot \cos\theta,  \quad v_\bot = v\cdot \sin\theta
\end{equation}

\begin{figure}[htb]
	\centering
	\includegraphics[width=0.28\textwidth]{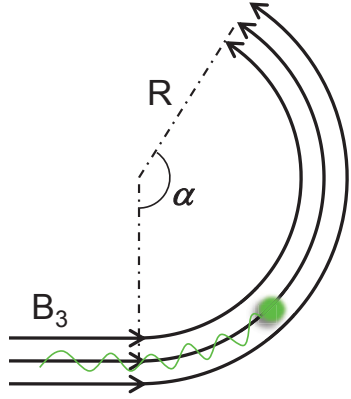}
	\caption{	Sketch of the principle of the \RB drift spectrometer, in which a uniformly curved magnetic field $B_3$ is generated. The magnetic field lines are distributed parallelly and coaxially, and are bended by an angle of $\alpha$.}
	\label{figure:RBspec concept}
\end{figure}

Suppose that we apply a uniformly curved magnetic field, with the curvature $R$ and the field strength $B$ as constants, then the curved magnetic field lines are distributed parallelly and coaxially, as shown in Fig.~\ref{figure:RBspec concept}.

According to Eq.~\ref{form:drift velocity}, the drift velocity $\mathbf{v}_d$ is a constant in the uniformly curved magnetic field, and the higher order contributions induced by $\dot{\mathbf{v}}_d$ are zero \cite{Erz09,Kru62,Lit83}. During a propagating time of $T$, the drift distance $D$ of a particle is

\begin{equation}
	D(p,\theta) = \int_T v_d \; \ud t = \frac{p}{qB}\cdot \alpha \cdot f(\theta)
	\label{form:drift distance}
\end{equation}

where $\alpha$ is the bending angle of the route of the particle gyration center during the time $T$

\begin{equation}
	\alpha = \frac{v_\parallel T}{R}
\end{equation}

as marked in Fig.~\ref{figure:RBspec concept}. 
$f(\theta)$ is a factor related to the particle pitch angle

\begin{equation}
	f(\theta)= \frac{1}{2}\left( \cos\theta+\frac{1}{\cos\theta} \right)
	\label{form:RB theta factor}
\end{equation}

Compare Eq.~\ref{form:drift distance} with Eq.~\ref{form:magspec disperse}, the behaviour of the \RB drift is similar as that of the particle dispersion in the magnetic spectrometer. The drifts or the dispersion distances in both cases are proportional to the particle momentum, and inversely proportional to the analysing magnetic field and the particle charge. 

\begin{figure}[htb]
	\centering
	\includegraphics[width=0.48\textwidth]{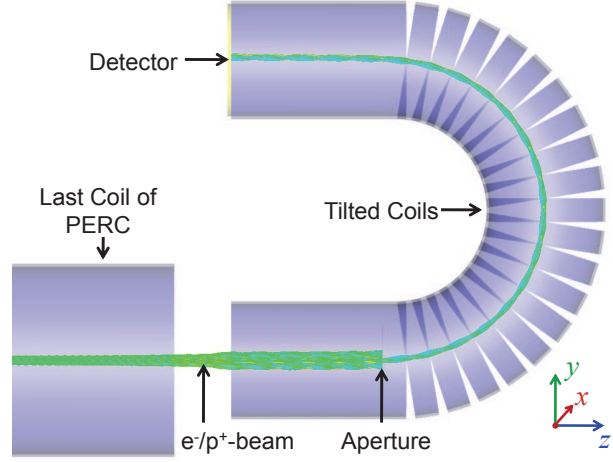}
	\caption{The design of the \RB drift spectrometer at the end of PERC, and the simulated trajectories of \ep.}
	\label{figure:driftspec geometry1}
\end{figure}

With this principle, we propose a design of the \RB drift momentum spectrometer, as shown in Fig.~\ref{figure:driftspec geometry1}.

\begin{figure}[htb]
	\centering
	\includegraphics[width = 0.5\textwidth]{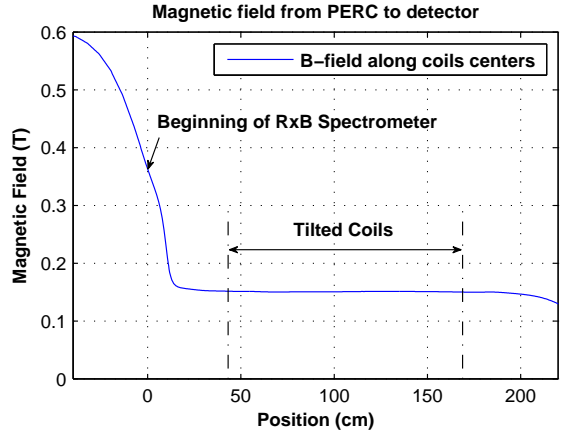}
	\caption{The magnetic field strength along the central line of the coils from the end of PERC to the detector of \RB spectrometer. The field is decreased gradually from $B_2$ = 0.5 T at the end of PERC to $B_3$ = 0.15 T, and is kept as constant in the tilted coils. The zero position denotes the beginning of the \RB spectrometer.
	}
	\label{figure:driftspec B-field}
\end{figure}

At the beginning of the \RB spectrometer, we apply several connected coils to gradually decrease the guiding field of PERC from 0.5 T to 0.15 T, with the field gradient satisfies the adiabatic condition of \ep transports. After that, a series of tilted coils generates a 180$\degree$ bended magnetic field. Along the central line of the tilted coils, the curvature of magnetic field line is $R_0$ = 40 cm, and the field strength is kept as $B_3$ = 0.15 T. Fig.~\ref{figure:driftspec B-field} plots the magnetic field from the end of PERC to the \RB spectrometer detector.

At the beginning of the tilted coils, we apply an aperture of 1\TT1 cm$^2$ to define the size of the incident \ep-beam. The particles through the aperture follow the curved magnetic lines, and turn 180$\degree$ then reach the detector on top. During the propagation, they drift along the $x$-axis according to their charges and momenta. 


Hence the \RB drift spectrometer, instead of eliminating the guiding field of PERC, evolves the field smoothly and gradually to the analysing magnetic field. The particles can be transported adiabatically during the processes, and their angular information can be kept and measured.

Table \ref{table:driftspec standard} lists the parameters of the standard configuration of the \RB spectrometer design. 

\begin{table}[htb]
\centering
\begin{tabular}{lll}
\toprule
Parameter	&	Comment	&	Value	\\
\midrule
$B_3$	&	Analysing field	&	0.15 T	\\
$R_0$	&	Field line curvature	&	40 cm	\\
$w$	&	Aperture width	&	1 cm	\\
$h$	&	Aperture height	&	1 cm\\
$\alpha$	&	Bending angle	&	$\pi$	\\
$\theta_{max}$	&	Max. pitch angle	&	9.1$\degree$	\\
$r_{max}$	&	Max. gyration radius	&	0.42 cm	\\
$D_{max}$	&	Max. drift	&	8.29 cm\\
\bottomrule
\end{tabular}
\caption{Parameters of the standard configuration of the \RB drift spectrometer.}
\label{table:driftspec standard}
\end{table}

$w$ and $h$ in Table \ref{table:driftspec standard} are the width and height of the aperture along the $x$- and $y$-axes. $\theta_{max}$ and $r_{max}$ are the maximum pitch angle and gyration radius of the decay \ep in the $B_3$ field 

\begin{equation}
	0 \leqslant \theta_3 \leqslant \theta_{max} = \arcsin \sqrt{\frac{B_3}{B_1}},\quad 0 \leqslant r \leqslant r_{max} = \left|\frac{p_{max}}{qB_3}\right|\cdot \sin \theta_{max}
\end{equation}

where $p_{max}$ = 1.19 MeV/c is the maximum momentum of \ep from the free neutron decay \cite{PDG10}. 

With the standard configuration, the maximum drift $D_{max}$ for $p_{max}$ is 8.29 cm.

\section{Corrections on \RB drift spectrometer}
\label{section:correction}

In the \RB spectrometer, the particle distribution on the detector is influenced by the properties of the particles and the instrument.

\subsection{\ep pitch angle $\theta$}

\begin{figure}[htb]
	\centering
	\includegraphics[width=0.5\textwidth]{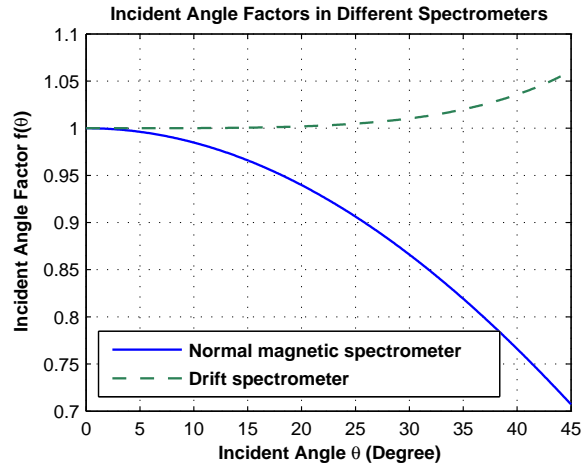}
	\caption{The incident angle factors $f(\theta)$ in the magnetic spectrometer and the \RB drift spectrometer, versus the incident angle $\theta$ from 0$\degree$ to 45$\degree$.}
	\label{figure:incident angle}
\end{figure}

Compare Eq.~\ref{form:RB theta factor} with Eq.~\ref{form:magspec disperse}, the influence of the correction factor $f(\theta)$ induced by the incident angle of particles in the \RB spectrometer is much smaller than that in the magnetic spectrometer. In Fig.~\ref{figure:incident angle}, the magnitudes of both correction factors are plotted.

For $\theta\leqslant$ 10$\degree$, the $f(\theta)$ in the \RB drift is negligible as less than 10$^{-4}$ deviated from 1, while the deviation in the magnetic spectrometer is 1.5\TT10$^{-2}$. Hence the \RB spectrometer has large acceptance of \ep incident angles, which is a significant advantage.

\subsection{Gyration radius $r$ and aperture width $w$}
	
In the \RB spectrometer, the gyration radii of the particles remain during the drifts and the detection. 
For a given momentum, the maximum gyration radius of the particles is

\begin{equation}
	r_c(p) = \left| \frac{p}{qB_3} \right| \cdot \sin\theta_{max}
\end{equation}

Through the aperture, 
the size of the \ep-beam on the detector will be 

\begin{equation}
	S_{\textrm{\ep-beam}} = W \times H = (w+ 4r_{c}) \times (h+4r_{c})
\end{equation}

as sketched in Fig.~\ref{figure:drift pattern} \textbf{(a)}. Therefore, the maximum deviation induced by the gyration radius $r_{c}$ relative to the drift $D$ is

\begin{equation}
	\frac{4 \, r_{c}(p)}{D(p,\theta_3)} = \frac{4}{\alpha} \frac{\sin\theta_{max}}{f(\theta_3)} \approx \frac{4}{\alpha}\cdot \sqrt{\frac{B_3}{B_1}}
\end{equation}

\subsection{Beam height $H$ and curvature $R$}

In the curved magnetic field, the field strength has a gradient along $\mathbf{R}$

\begin{equation}
	\nabla\times \mathbf{B} = 0, \quad B_{3}\cdot R_{0} = B\cdot R
\end{equation}

$B_{3}$ and $R_{0}$ are the field strength and the curvature along the central line of the tilted coils. The particles at different positions along $\mathbf{R}$ will experience deviated magnetic fields, thus have the drift

\begin{equation}
	D=\frac{p}{qB_{3}}\cdot \alpha \cdot f(\theta_3) \cdot \frac{R}{R_{0}} = D_0 \frac{R}{R_{0}} = D_0 \frac{R_0+y}{R_0}
	\label{form:RB yh correct}
\end{equation}

where $y$ is the vertical position of the particle on the detector relative to the beam center. The maximum deviation of the drift is

\begin{equation}
	\frac{\Delta D}{D_0} = \frac{h+4r_c}{R_{0}} = \frac{H}{R_0}
\end{equation}

Hence the \ep-beam height relative to the field curvature $H/R_0$ tilts the particle distribution on the detector, as sketched in Fig.~\ref{figure:drift pattern} \textbf{(b)}. For a detector only sensitive to the $x$-position, the measured particle distribution is widened.

\subsection{Particle distribution on detector}

\begin{figure}[htb]
	\centering
	\subfloat[Electron distribution influenced by $w$ and $r$.]{
	\includegraphics[width=0.45\textwidth]{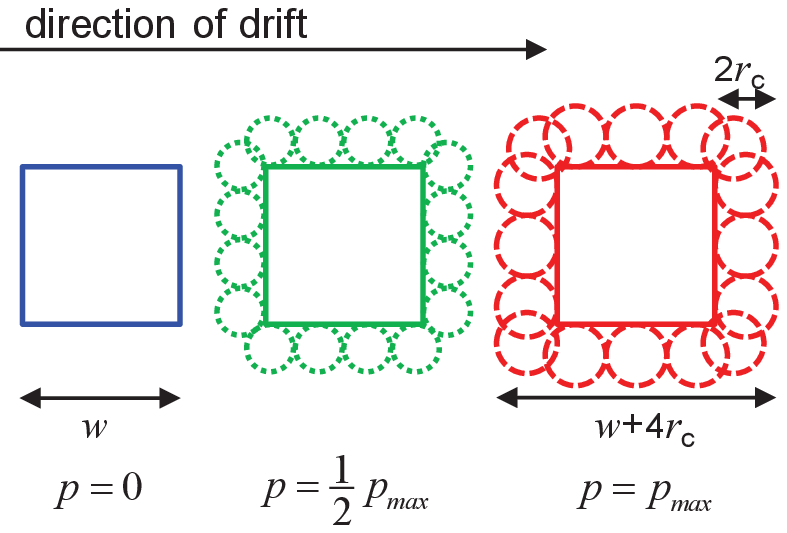}
	}\hspace{5pt}
	\subfloat[Electron distribution influenced by $H$.]{
	\includegraphics[width=0.45\textwidth]{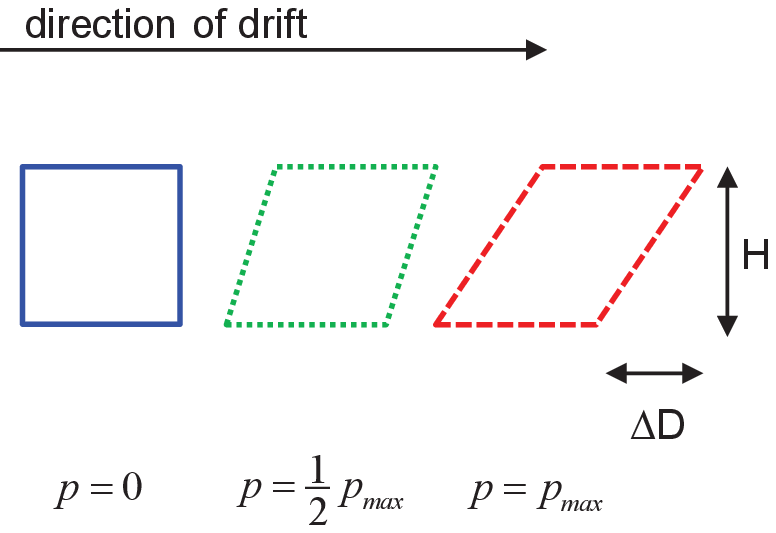}
	}
	\caption{Sketches of the electron pattern on the detector with discrete momenta, influenced by $w$ , $r$, and $H$. $p_{max}$ denotes the maximum electron momentum in free neutron decay.}
	\label{figure:drift pattern}
\end{figure}

The dispersion of the particles in the drift spectrometer is mainly influenced by the systematics of $w$, $r$, and $H$. Fig.~\ref{figure:drift pattern} shows the sketchy distribution of electrons on the detector affected by these factors. 

\begin{figure}[htb]
	\centering
	\includegraphics[width=\textwidth]{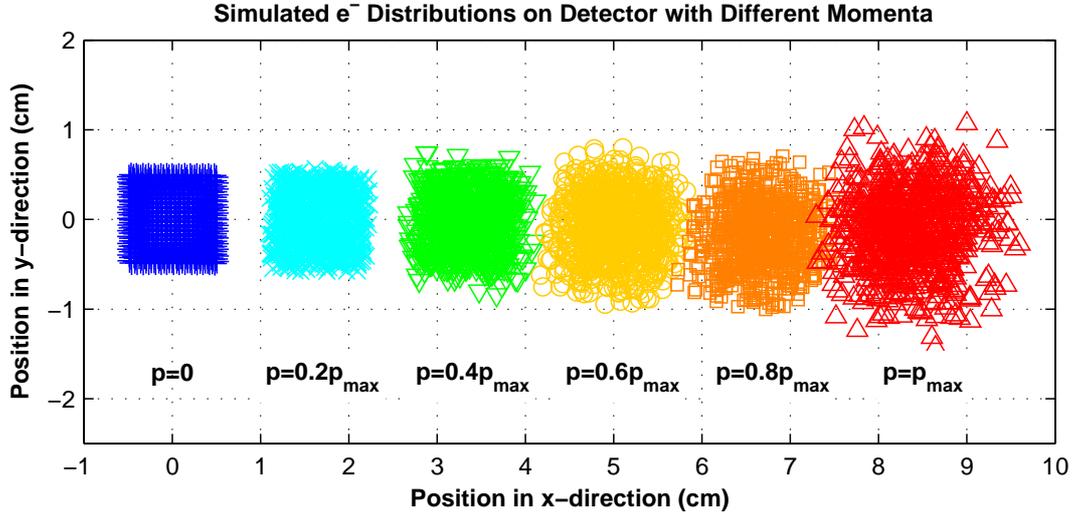}
	\caption{Simulated electron distribution on the detector of the \RB spectrometer with discrete momenta. $p_{max}$ denotes the maximum electron momentum in free neutron decay as 1.19 MeV/c. The particles with $p\rightarrow 0$ can also be measured, and the deviations caused by $r$ and $H$ equal zero.}
	\label{figure:drift scatter}
\end{figure}

Fig.~\ref{figure:drift scatter} shows the simulated distributions of electrons on the detector, with discrete momenta from 0 to 1.19 MeV/c.

The deviation of the drift caused by $w$ is a constant, while that induced by $r$ and $H$ are proportional to the drift distance. Hence at the low momentum range, the \RB spectrometer has better performance.

Furthermore, in normal magnetic spectrometer, the particles with $p\rightarrow 0$ cannot be totally measured if their dispersion distances are smaller than the aperture width $D<w$. While as shown in Fig.~\ref{figure:drift scatter}, the \RB spectrometer can measure the full range of the momentum.

\section{Transfer function}
\label{section:TF}

The motions of the \ep particles in the \RB spectrometer are clearly defined during the drift processes. The transfer function, i.e., the relation between the momentum spectrum $F(p)$ and the particle distribution on the detector $G(x)$, can be calculated and used in the data analyses. In this section, we discuss the transfer function including the corrections of $r$, $w$, and $H$. The incident angle correction $f(\theta)$ and the higher order contributions, e.g., the $R$ deviation induced by $v_d$, the $r$ deviation induced by the $B_3$ gradient along $R$, are not considered here.

\subsection{Particle distribution from point source}

We assume the \ep from the decay volume of PERC are homogeneously distributed in the open window of the aperture. If the aperture is sufficiently thin, it does not distort the angular distribution of the particles. In addition, the size of the \ep-beam in front of the aperture is much larger than that of the aperture. Therefore, the particles that pass through the aperture, can be treated as emitted from the aperture's open window.

For the particles emitted from a point source at position $x=0$ with given momentum and pitch angle, their distribution along the $x$-axis is given as \cite{Dub08PSF}

\begin{equation}
	 g(x,p,\theta_3) = \frac{1}{\pi^2 r(p,\theta_3)} K \left(1-\frac{x^2}{4 r^2(p,\theta_3)} \right)
	 \label{form:PSF}
\end{equation}

where $K$ denotes the complete elliptical integral of the first kind.

\subsection{Angular distribution of particles in $B_3$ field}

If we apply unpolarized neutrons in PERC experiments, the \ep are isotropically emitted in the decay volume \cite{Jac57}. Their angular distribution in the field $B_0$ is

\begin{equation}
	W_0(\theta_{0}) = \frac{\ud N}{\ud \theta_{0}} = \frac{1}{2} \sin \theta_{0}
\end{equation}

With Eq.~\ref{form:adia theta}, when the particles propagate from $B_0$ to $B_3$, their angular distribution is

\begin{equation}
	W(\theta_3) = \frac{\ud N}{\ud \theta_3} = \frac{1}{2} \frac{B_0}{B_3} \frac{\sin\theta_3 \cos\theta_3}{\sqrt{1 - \frac{B_0}{B_3} \sin^2\theta_3}}\quad \mathrm{with} \quad 0 \leqslant \theta_3 \leqslant \theta_{max}
\end{equation}

We integrate Eq.~\ref{form:PSF} over $\theta_3$, the particle distribution along the $x$-axis from a point source is then

\begin{equation}
	L(x,p) = \int^{\theta_{max}}_0 W(\theta_3) \cdot  g(x,p,\theta_3) \, \ud \theta_3
	\label{form:RB Lx}
\end{equation}

\begin{figure}[ht]
	\centering
	\includegraphics[width=0.5\textwidth]{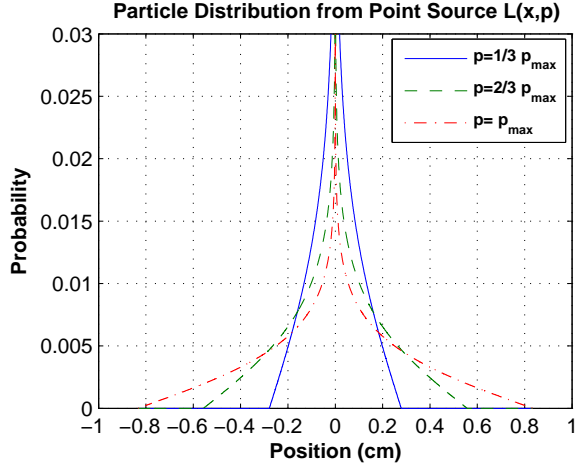}
	\caption{	The distribution $L(x,p)$ of particles from a point source in the $B_3$ field along the $x$-axis, as given in Eq.~\ref{form:RB Lx}. Different curves denote the particles with different momenta. $p_{max}$ is the maximum momentum of \ep in free neutron decay as 1.19 MeV/c.}
	\label{figure:RB Lx}
\end{figure}

Fig.~\ref{figure:RB Lx} plots the distribution $L(x,p)$ of particles in the $B_3$ field with different momenta.

\subsection{Transmission function of aperture}

Since the distance from the aperture to the detector is much longer than the helical pitches of the particles, we assume the particle distribution on the detector from any point in the aperture follows Eq.~\ref{form:RB Lx}. 

Define the transmission function of the aperture along the $x$-axis as

\begin{equation}
	T(x)  = \begin{cases}
	 1\; ;	\qquad\quad &  -w/2 \leqslant x \leqslant w/2 \\
	 0\; ; & x < -w/2\; , \;  x > w/2
	\end{cases}
	\label{form:RB TF2}
\end{equation}

On the detector, the particles from the open window of the aperture then have a distribution of

\begin{equation}
	P(x,p) = T(x) \ast L(x,p)
	\label{form:RB TF xw}
\end{equation}

which is the convolution of the aperture function $T(x)$ and the distribution $L(x,p)$ of point source.

\subsection{Correction of $H$}

For the correction of beam height $H$, we consider the particle distribution along the $y$-axis. For simplicity, we use the condition $w$ = $h$, so the distribution along the $y$-axis is the same as Eq.~\ref{form:RB TF xw}, i.e., $P(y,p)$. Since the factor $H/R_0$ tilts the particle distribution on detector, $P(y,p)$ is projected on the $x$-axis. According to Eq.~\ref{form:RB yh correct}, the projection can be written as

\begin{equation}
	Q(x,p) = \frac{R_0}{D(p)} \cdot P \left( \frac{x\cdot R_0}{D(p)},p \right) 
	\label{form:RB TF yh}
\end{equation}

\subsection{Transfer function}

All together, we take the corrections related to $r$, $w$, $H$, as well as the drift $D(p)$ into account. The total transfer function is then the convolution of the distributions in Eq.~\ref{form:RB TF xw} and Eq.~\ref{form:RB TF yh}

\begin{equation}
	M(x,p) =  [P( p ) \ast Q( p ) ](x-D(p))
\end{equation}

For a given momentum spectrum $F(p)$, the particle distribution on the detector can be derived

\begin{equation}
	G(x) = \int_0^{p_{max}} M(x,p) \cdot F(p) \; \ud p
	\label{form:RB Fredholm}
\end{equation}

\begin{figure}[ht]
	\hspace{-5pt}	\vspace{0pt}
	\subfloat[$F(p)$]{
	\includegraphics[scale=1]{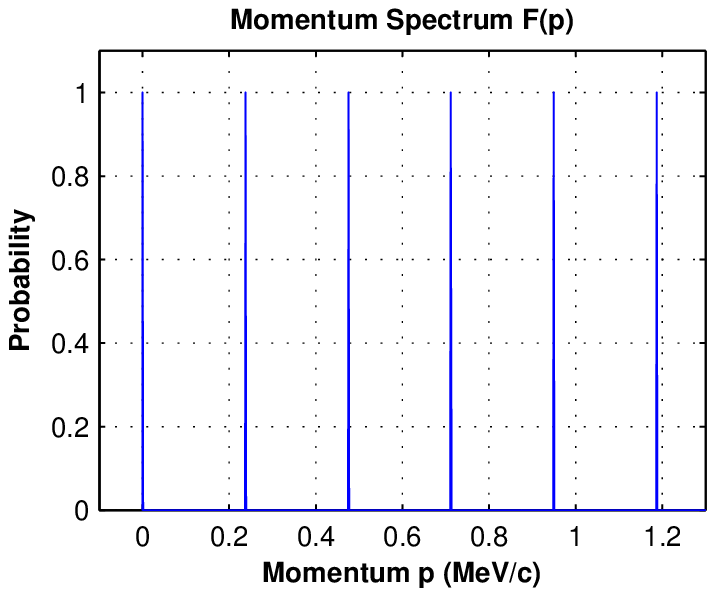}
	}\hspace{-15pt}
	\subfloat[$G(x)$]{	
	\includegraphics[scale=1]{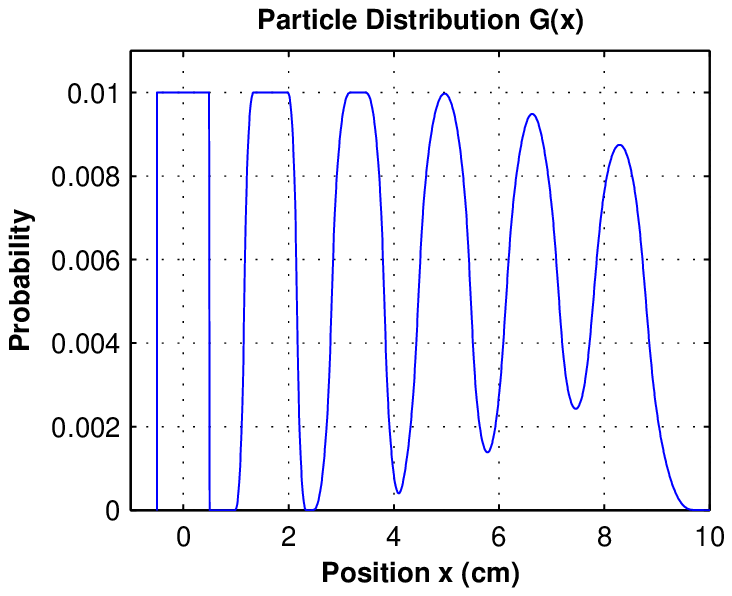}
	}\\ 
	\hspace{-0pt}\vspace{0pt}
	\subfloat[$F(p)$]{
	\includegraphics[scale=1]{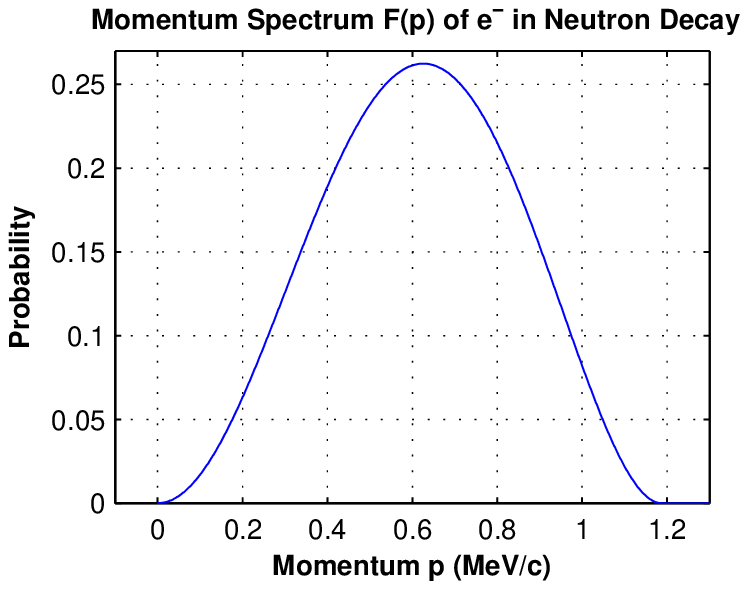}
	}\hspace{-15pt}
	\subfloat[$G(x)$]{
	\includegraphics[scale=1]{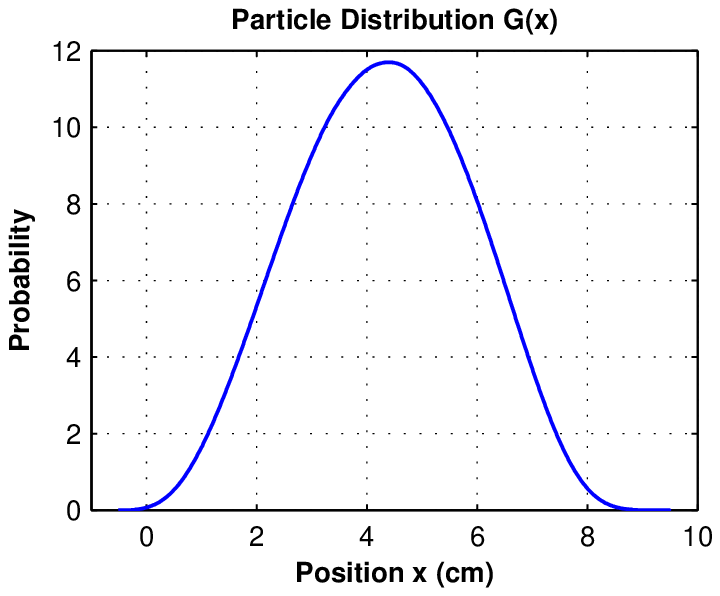}
	}
	\caption{Examples of given $F(p)$ and resulted $G(x)$ according to Eq.~\ref{form:RB Fredholm}. Fig.~\textbf{(a)} and \textbf{(b)} denote discrete momentum and position distribution same as shown in Fig.~\ref{figure:drift scatter}. Fig.~\textbf{(c)} and \textbf{(d)} denote the theoretical momentum spectrum of electrons in free neutron decay and their position distribution.}
	\label{figure:RB TF eg}
\end{figure}

Fig.~\ref{figure:RB TF eg} shows the examples of the given momentum spectra $F(p)$ and the resulted $G(x)$ from Eq.~\ref{form:RB Fredholm}.

In experiments, $G(x)$ can be measured by position sensitive detectors. To evaluate $F(p)$ from Eq.~\ref{form:RB Fredholm}, which is the Fredholm integral equation of first type, a numerical method is to convert the integral into the quadrature calculation, the $F(p)$ and $G(x)$ into arrays, thus convert the kernel function $M(x,p)$ into a square matrix \cite{Del88,Pre07}. The transfer function then can be written as

\begin{equation}
	\mathbf{G} = \mathbf{M \cdot F}
\end{equation}

and the momentum spectrum $\mathbf{F}$ can be evaluated

\begin{equation}
	\mathbf{F} = \mathbf{M^{-1} \cdot G}
	\label{form:RB inverse K}
\end{equation}

In this case, the $\mathbf{G}$ and $\mathbf{F}$ have the same number of rows, thus the same resolution. In the standard configuration, if $\mathbf{G}$ has a resolution of 1 mm, the momentum spectrum $\mathbf{F}$ can reach a resolution of 14.4 keV/c.

\section{Conclusion}

The \RB drift spectrometer offers the opportunity of momentum measurements of charged particles in the instrument with guiding fields, in which case the normal magnetic spectrometers cannot work well. In this proposed drift spectrometer, the guiding field is not eliminated, but gradually evolved to the analysing field. The drifts of the particles in the uniformly curved magnetic field have similar behaviours as in the normal magnetic spectrometer.

As a conclusion, the \RB spectrometer has the advantages:
\begin{itemize}
	\item \emph{Adiabatic transports of particles.} 
	As shown in Fig.~\ref{figure:driftspec B-field}, from the guiding field to the detector of the \RB spectrometer, the charged particles can be adiabatically transported. The angular distribution of the particles can be kept and measured.
	\item \emph{Low momentum measurements.} 
	As shown in Fig.~\ref{figure:drift scatter}, the particles with very small momentum $p\rightarrow 0$ can be measured in the \RB spectrometer, while they cannot be measured in normal magnetic spectrometer if their dispersion $D<w$.
	\item \emph{Large acceptance of incident angle.} 	
	As shown in Fig.~\ref{figure:incident angle}, the direct correction induced by the \ep incident angle is very small as less than 10$^{-4}$ when $\theta\leqslant$ 10$\degree$.
\end{itemize}

As a conceptual design of \RB spectrometer, the particle drifts are considerably influenced by the systematics related to both the instrument and the particle properties. Table \ref{table:driftspec correction} lists the maximum sizes of the corrections in the standard configuration. 

\begin{table}[htb]
\centering
\begin{tabular}{lcc}
\bottomrule
Correction	&	Comment & Max.~Size \\
\hline
4$r_{c}/D$	&	Gyration Radius &	2.0\TT10$^{-1}$\\
$w/D_{max}$		&	Aperture Width &	1.2\TT10$^{-1}$\\
$H/R_0$	&	Aperture Height &	6.7\TT10$^{-2}$\\
$\Delta B_3$	& Field Homogeneity	&	8\TT10$^{-3}$\\
$f(\theta_{max})$	&	Incident Angle &	8\TT10$^{-5}$\\
\toprule
\end{tabular}
\caption{Estimated maximum correction sizes on the \ep particles dispersion in the \RB spectrometer.}
\label{table:driftspec correction}
\end{table}

However, the motions of the \ep particles are clearly defined during the drifts, thus the transfer function of the particles can be well known. Experimentally, one is able to fit the momentum spectrum $F(p)$ to the measured particle distribution $G(x)$, or numerically evaluate $F(p)$ from $G(x)$. In the standard configuration, if the position detector has a resolution of 1 mm, the momentum spectra can reach a resolution of 14.4 keV/c. Additionally, by performing detector calibration with defined particle sources, the systematic errors can be controlled at low level.

Besides, there is also room for improvement of the \RB spectrometer design. For further development, we can decrease the corrections, e.g., by optimizing the bending angle $\alpha$, the curvature $R$, and the analysing magnetic field $B_3$. The transfer function in this article only considers the motions of the \ep particles to the first order. The higher order contributions, e.g., the acceleration $\dot{\mathbf{v}}_d$, the deviation of curvature $\mathbf{R}$ induced by the drift $D$, are more complicated \cite{Kru62,Lit83}. To obtain the transfer function with high accuracies, we can apply fine simulations of \ep trajectories in the \RB spectrometer.

\section{Acknowledgement}

We would like to thank MSc.~Z.~Lee (Electron Microscopy Group of Materials Science, Ulm University) for helpful discussions about the transfer function. This work is supported by the German Research Foundation as part of the Priority Program 1491, and the Austrian Science Fund under contracts No. I 528-N20 and I 534-N20.

\bibliography{reference}

\end{document}